# ARCHAEOASTRONOMY AND THE ORIENTATION OF CHURCHES IN THE JESUIT MISSIONS OF NORTH-WESTERN NEW SPAIN

María Florencia Muratore[1,2], Adrián Di Paolo[3] and Alejandro Gangui[1,3]

[1] CONICET - Universidad de Buenos Aires, Instituto de Astronomía y Física del Espacio (IAFE), Argentina
[2] Universidad Nacional de Luján, Departamento de Ciencias Básicas, Buenos Aires, Argentina
[3] Universidad de Buenos Aires, Facultad de Ciencias Exactas y Naturales, Argentina
flormuratore@gmail.com, adriandipaolo@gmail.com, gangui@iafe.uba.ar

**Abstract:** We present an extensive archaeoastronomical study of the orientations of seventeenth- and eighteenth-century Jesuit churches in the lands of the historic viceroyalty of New Spain. Our sample includes forty-one chapels and churches located mainly in present-day Mexico, which documentary sources indicate were built by the Society, and for which we measured the azimuths and heights of the horizon of their principal axes using satellite imagery and digital elevation models. Our results show that neither the orientation diagram nor the statistical analysis derived from the sample declination histogram can select a particular orientation pattern with an adequate level of confidence. We suggest some possible explanations for our results, discussing these North American churches within a broader cultural and geographical context that includes previous studies involving Jesuit mission churches in South America. Based on the analysis of the data presented here, we conclude that the orientation of Jesuit churches in the viceroyalty of New Spain most likely does not follow a well-defined prescription.

**Keywords:** Christian churches, archaeoastronomy, orientations, Society of Jesus, Viceroyalty of New Spain

## Introduction

The study of the orientation of Christian churches has been of interest for a long time, having received new impetus in the literature as it was recognized that it represents a key feature of their architecture (González-García and Belmonte, 2015). Based on the texts of the early Christian writers, churches' apses should lie along a particular direction, that is, the priest had to stand facing eastward during services. This is recognized by Clement of Alexandria, Tertullian and Origen, and it might have been formalized during the first Council of Nicaea (AD 325). Also, in the fourth century St. Athanasius of Alexandria declared that the priest and participants should face east, where Christ, the Sun of Justice, would shine at the end of time (McCluskey, 2015).

There have been many works that have concentrated mainly on studying groups of churches belonging to a particular historical period or to a particular architectural style. In our previous works we have also concentrated on the study of groups of

113

colonial churches distributed over a limited region, as it happens in the insular territory of the Canary Islands or a certain region of the highlands (*altiplano*) in northern Chile (Gangui et al., 2016). These studies generally considered churches of various religious orders together and provided a homogeneous and culturally significant case study. But it is also interesting to study the religious constructions of a particular Order in a limited period of time.

In this work we focus on the Jesuit missionary churches in America, which for almost two centuries were the most representative constructions in the process of Christian evangelization on the continent until the Order's expulsion from Spanish-ruled territory in 1767. The main objective is to discern possible patterns of orientations in the studied structures and to assess whether these orientations could be related to the location of the Sun or other celestial bodies when crossing the local horizon, which could yield important information pertaining to their construction. In the following sections we succinctly present the historical and cultural elements characterizing the Society of Jesus and their missions and churches in South America, and then focus on new archaeoastronomical work on the orientation of churches in lands of present-day Mexico and surrounding regions.

### The Society of Jesus

As a response of the Catholic Church to the Protestant Reformation, the Jesuits (members of the Society of Jesus) were recognised as a new religious order by Pope Paul III in 1540. Founded by the Basque priest Ignatius of Loyola, "The Society" (also known as *La Compañía*) became the spiritual and intellectual guide of Catholic Europe for the next two and a half centuries. Through arduous and extensive travels abroad, their mission was to spread Christianity and save souls, and to propagate the faith in unconverted lands. In the interim, they established a cohesive worldwide network that attracted prominent and wealthy individuals. Jesuit dominance brought power to the organisation and influence in all walks of life, as well as political favours, important financial contributions, and somewhat inordinate freedoms. Spain's Habsburg rulers favoured the Society because they volunteered for the toughest challenges and kept their promises, earning the distrust and envy of their religious competitors. One of those tough tasks, of course, was the evangelisation of natives in the newest lands conquered by the Crown, for example, in the New World. This favouritism began to wane when the Bourbon rulers of Spain came to power in the 1700s, with their strict control of finances and their suspicion of Jesuit activities and commercial operations. Little more than half a century later the Society would be expelled from America and then dissolved.

### Jesuit missions in South America

We may say, quoting the historian of science Miguel de Asúa, that *Paraquaria* (the historical Jesuit Province of Paraguay, which also included the northeast of present-day Argentina and southern Brazil), and also *Chiquitos* (the Jesuit Province in present-day eastern Bolivia), have been seen as a mirage of Baroque splendor in the middle of the luxuriant vegetation of the tropical forest (Asúa, 2014). The reductions (*reducciones*) of South America, that is, settlements for indigenous people organized and administered by Jesuit priests, particularly of historical Paraguay and Chiquitos, were the living core of the complex of the religious, economic, and educational institutions which the Society of Jesus built up in the continent.

The Fathers arrived in Paraquaria in the 1580s, and more than a century afterwards in Chiquitos (Querejazu, 1995), but their mission did not bear tangible results until several decades later. The reductions allowed the Jesuits to segregate both the Guaraní and the Chiquitanos from Spanish colonial society, and to protect them from the way of life of the settlers -who wished to take them as slaves-, while instructing them in the Christian doctrine. Under the rigid and disciplined organization imposed by the missionaries the Indians in these villages found security for their families and material well-being, but at the price of absolute dedication to Roman Catholicism and a complete change in their old ways of life.

Keeping with the planned religious-centered nature of the Jesuit missions, their layouts were not left to chance. The organization of the urban space in the missionary villages was centered on the main square (*la plaza*), and on one side of it included the church as the main place of worship, the college-residence and the cemetery. These constructions, together with the main Jesuit courtyard and the bell tower, formed the essential religious nucleus in the configuration of the space. Indian houses covered the other three sides of the main square.

Detailed studies of the orientation of Jesuit churches have been carried out recently in both extended regions of Paraquaria (Giménez Benítez et al., 2018) and Chiquitos (Gangui, 2020). From these studies it could be shown that, unlike the Paraquaria churches where the orientations are mostly meridian in the north-south direction, half of the studied churches in Chiquitos have canonical orientations that appear to be aligned with solar phenomena, with their altars oriented towards astronomical declinations that fall within the solar range, and with three notable churches exhibiting a precise equinoctial orientation, as shown in Fig. 1 (Gangui, 2020).

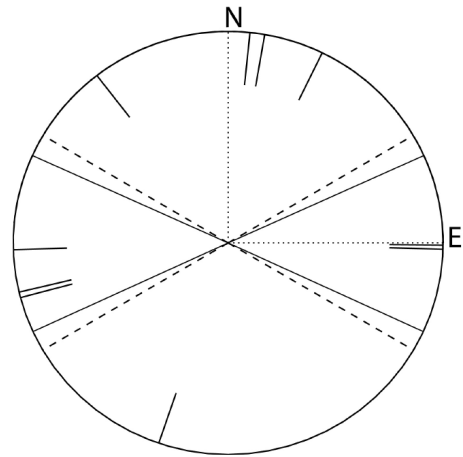

Fig. 1. Orientation diagram of the Jesuit churches in the province of Chiquitos (Gangui, 2020).

In this project, in the light of our previous work, we try to understand whether it was the new landscape of the South American virgin forest what made the missionaries deviate from the indications of the old Catholic writers, as is already partly the case with the colonial churches in some of the Canary Islands (e.g., Di Paolo et al., 2020), or whether, on the contrary, there was an initial intention in the Jesuit custom to orientate their monumental churches (their particular way of building, or "modo nostro", cf.,

Garofalo (2015)) that could justify the data that were obtained at these sites.

The adaptability of the Jesuits to diverse circumstances is well documented, as is the perfect balance recommended to them for the configuration of all activities or policies, and this could also include the design and construction of their churches. To extend our knowledge of the Order to other regions where they carried out their evangelization activity, we will now concentrate on the missions of North America, in the viceroyalty of New Spain.

### Jesuit churches in the Viceroyalty of New Spain

By 1572, a few decades after the founding of the Order, Jesuit priests arrived and settled in Central America, as did members of the Dominican and Franciscan mendicant orders. In the area of interest for our study, permanent missions began to be established in the Sonoran Desert from about 1640, and later, in 1697, the first post on the Baja California peninsula to survive to the present day, the mission of *Nuestra Señora de Loreto Conchó*, was founded (Fig. 2). Over time, Loreto became the head of the peninsula's missions and extended its influence into the surrounding regions. From these strategic sites, both on the peninsula and on the mainland, the Jesuits attracted nomadic Native Americans and grouped them into missions for baptism and conversion. For more than a hundred years they organised the Indians to live in society and cultivate the land. And the Indians, as was the case in other missions on the continent, were instrumental in helping to build the solid, ornate stone churches that populated the region.

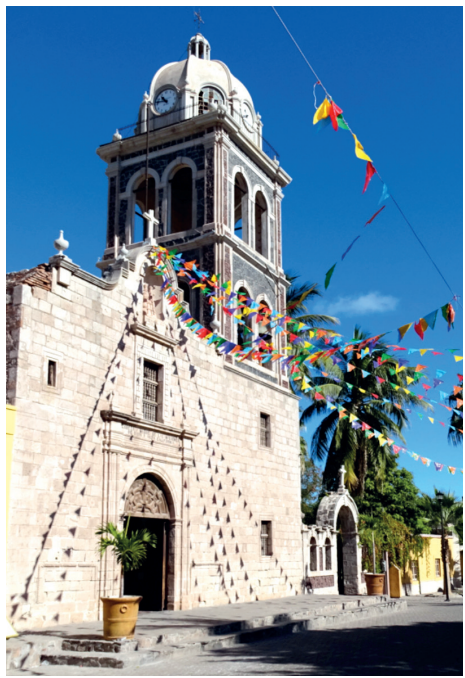

Fig. 2. The church Our Lady of Loreto, in the present city of Loreto, Baja California Sur, Mexico. Photograph by Adam Jones (Wikipedia Commons).

The results of the studies already developed in South America (and summarised in the previous section) must now be interpreted within a broader historical and geographical context. For we know that before the churches studied in Chiquitos, for example, Jesuits built for decades many churches on the other side of the equator (as already noted, in Sonora and in neighbouring states of present-day Mexico). Therefore, to obtain a more complete picture of religious architecture on the continent, it is necessary to compare the data already analysed with the orientation of Jesuit churches of the 17th and 18th centuries in the former viceroyalty of New Spain in North America.

**Data selection**

We conducted a detailed exploration of the present-day Mexican states of Baja California, Sonora, Chihuahua, Sinaloa, Durango, and other nearby regions, in search of standing Jesuit churches or ruins of old temples that are properly documented and have a distinguishable structure. We used satellite maps together with various documentary sources and selected a large number of existing churches in the area that were built by *La Compañía* in the seventeenth and eighteenth centuries and for which it was possible to measure their spatial orientations (Fig. 3). By using the maps and the measuring tools of Google Earth we were able to estimate, with an accuracy of a few degrees, the actual azimuths of the main axes of the churches.

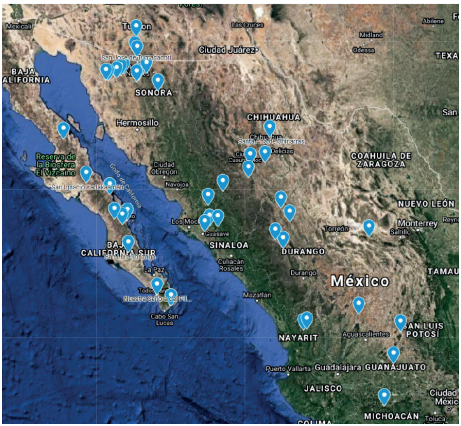

Fig. 3. Group of Jesuit churches from the 17th and 18th centuries with azimuth measurements, located in territories that belonged to the viceroyalty of New Spain.

The measured and calculated data for each of the churches in the sample are presented in Table 1 (at the end of the manuscript). The penultimate column of the table includes the angular height of the point of the horizon towards which the altar of the church is facing (h), corrected for atmospheric refraction. For estimating these values, we employed a digital elevation model based on the Shuttle Radar Topographic Mission (SRTM) available at HeyWhatsThat (Kosowsky 2017), which gives angular heights within a 0.5° approximation. When the horizon was too close, making uncertain the value of h, the Elevation Profile Tool of Google Earth was also employed to get a more precise value of the height of the horizon.

Fig. 4 shows the orientation diagram for the analyzed churches. The diagonal lines of the graph indicate the azimuths corresponding -in the eastern horizon- to the extreme values for the Sun (azimuths of 63.0° and 116.3° -shown in continuous lines-, equivalent to the northern summer and winter solstices, respectively) and for the Moon (azimuths of 57.0° and 123.2° -dashed lines-, equivalent to the position of the major lunistices or lunar standstills).

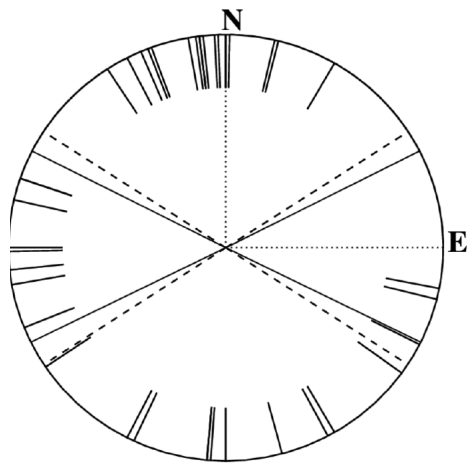

Fig. 4. Orientation diagram of the Jesuit churches in the the viceroyalty of New Spain.

This diagram shows that most of the altars in these churches are placed in such a way that the construction deviates from the canonical orientation *ad orientem*.

## Methodology

To better understand what we have discussed so far, here we present an analysis that employs all the data available, including the angular height of the horizon, h, and is based on the computation of the declination histogram. This will provide an estimation of the probability of obtaining particular declination values for a given sample of measurements.

In our analysis, we are using an appropriate smoothing of the declination histogram by a function called "kernel" to generate the kernel density estimate (KDE hereafter). For each entry in declination in Table 1, we multiply the value of the number of occurrences by the kernel function with a given passband or bandwidth. We have employed an Epanechnikov kernel with a width of roughly twice our estimated error in declination. Following González-García and Šprajc (2016, 192), to ensure that a concentration of values for the declination is significant, we must compare the distribution of our measured data against the result expected from a uniform distribution where the orientations are homogeneously distributed to each possible point in the horizon (the null hypothesis). This comparison quantifies the significance of our results.

Obtaining a KDE-smoothed histogram (a curvigram) scaled with respect to the uniform distribution allows us to see whether our actual data departs significantly from that distribution. As the scale is given by the standard deviation of such uniform distribution, if our data has a maximum that reaches the value 3, for example, it means that it is three times higher than that standard deviation, or $3\sigma$. We take this as the standard criterion to indicate whether our obtained values in declination are significant or not.

## Results and discussion

Our results (Fig. 4) show that a large majority of the Jesuit churches that remain standing in the historical region of New Spain are oriented towards the northern and western quadrants. Surprisingly few churches show orientations within the solar range towards the eastern sector of the horizon. This differs from other groups of colonial Christian temples (including Jesuit groups, such as some equinoctially orientated ones that were measured in Chiquitos) and, for the moment, has no explanation. We have reviewed the literature related to the architecture of *La Compañía* and, unfortunately, the sources so far consulted that describe some aspects of Jesuit constructions do not emphasize the spatial orientation of their historic churches (Montes González, 2011; Alcalá, 2012; Cuesta Hernández, 2013).

In Fig. 5 we present the declination histogram, which takes into account also the height of the horizon corresponding to measurements taken towards the altars. Thus, the resulting data is independent of geographical location and local topography. This figure shows the astronomical declination versus the normalized relative frequency, which enables a clear and more accurate determination of the structure of the peaks.

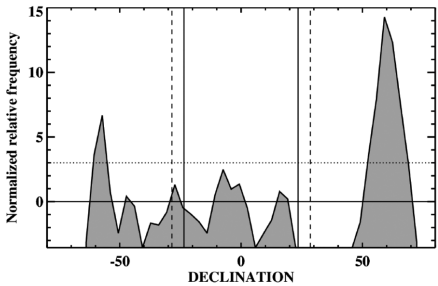

Fig. 5. Declination histogram (or curvigram), corresponding to the measurements taken along the axes of the churches towards the altars, showing the estimated "normalised relative frequency", as detailed in the main text. There are just two peaks above the 3σ level (dotted horizontal line). The continuous vertical lines represent the declinations corresponding to the extreme positions of the Sun at the solstices, while the dashed vertical lines represent the same for the Moon in major lunar standstills.

Note that above the 3σ level, indicated with a dotted horizontal line, there is no clearly defined concentration of declination values. Just a couple of peaks appear, with the most prominent one located beyond declination 50°, which most probably corresponds to an accumulation peak due to orientations near the meridian line (as is also suggested in Fig. 4). The other small peaks with declination values roughly in the range (-50°, +50°) are not statistically significant.

## Conclusions and future work

*I*n this study we measured and analysed the spatial orientation of forty-one Jesuit churches located in the missions of the viceroyalty on New Spain. Our main aim was to look for a possible pattern in the orientation of these outstanding buildings and to see whether it could be correlated with the changing positions of sunrises and sunsets throughout the year. We also provided our work with an adequate contextualization by means of a succinct cultural and historical study of the presence of the Society in the continent, both in South America and on the other side of the equator, in an extended region which today covers mainly present-day Mexico and Arizona.

We first reviewed previous archaeoastronomical studies conducted in the historical regions of Paraquaria and Chiquitos to emphasise that already two relatively numerous groups of colonial Jesuit churches did not respect, in general, the standard canonical pattern of orientations found in European medieval churches. The reasons for this are still uncertain, but it could be that the Jesuits had to adapt their missions to a brand-new landscape –the South American rainforest– never imagined before. And this is perhaps why the importance of the churches' orientations was put on the back burner.

The conjecture that the influence of the Jesuit's way of construction –the so-called "modo nostro"– could have been the responsible for the orientation pattern found, is a possible explanation (Gangui, 2020). This "way", which paid special attention to community needs and to the practical functionality of spaces for liturgy and living, was a natural consequence of the Order's flexibility to adapt to the particularities of each new territory.

We think that something similar applies to the new results we present in this paper. The orientation of the seventeenth- and eighteenth- century Jesuit churches we measured in present-day Mexico do not follow a recognisable pattern. In fact, the axes of these constructions orientate to roughly all sectors of the horizon, except notably to the eastern sector and within the solar range, as shown in Fig. 4. A more complete analysis including the angular height of the horizon, which leads to the peaks of the declination histogram of Fig. 5, shows that, apart from the extreme accumulation peaks located at high absolute values, the rest of the declination values are not statistically significant. This result prevents us from pointing to a particular direction –astronomical or otherwise– that is salient from the data.

It should be noted, however, that this study is not yet complete. In particular, we must delve more into the local cultural context regarding traditional sacred orientations, a project which will undoubtedly reveal new clues about the construction of the churches. We are also still surveying the territory and completing our database of churches, especially those located in the regions where the Jesuit occupation came later. We are currently working in an exploration of the region of Guanajuato and of the few remaining Jesuit missions in the Sierra of Nayarit, which represents the Society's last foundation project in New Spain and that of shortest duration (Gutiérrez Arriola, 2012).

Hence, based on the analysis of the data presented here, we could conclude that the orientation of the Jesuit churches of the viceroyalty of New Spain (and of the missions themselves, if we assume the church, the *plaza* and the rest of the buildings in the religious complex shared always a common axis with the small village, as is the case in Paraquaria and Chiquitos) most likely does not follow a well-defined prescription. Either the orientation was not a detail the Jesuit fathers consider important to focus on before the layout of each site, or they decided that every church should be adapted to each individual location and landscape.

Be that as it may, the results presented in this work provide a good basis for further exploration of the Jesuit churches of New Spain and complete a first survey of the churches that the Society built in America during its brief and unique evangelisation activity.

### Acknowledgements

We would like to thank Miguel de Asúa and César González-García for enlightening discussions on these topics and the constructive comments from two anonymous reviewers. This work was partially supported by CONICET and by the University of Buenos Aires. M.F.M is a doctoral fellow of CONICET.

Table 1. Orientations of the Jesuit churches in the provinces belonging to the viceroyalty of New Spain. L= latitude; l = longitude; a = astronomical azimuth; h = height of the horizon; δ = astronomical declination corresponding to the central point of the solar disc. When available, dates of the documented first foundation of the missions hosting their churches are included. These churches may have been built several decades later. BC stands for Baja California, and similarly for Sonora (So), Arizona (Ar), Chihuahua (Ch), Sinaloa (Si), Durango (Du), Michoacan (Mi), Nayarit (Na) and Coahuila de Zaragoza (Co)

| Location | Name | L (°) | l (°) | a (°) | h (°) | δ (°) |
|---|---|---|---|---|---|---|
| Loreto (BC) | Ntra. Sra. de Loreto (1697) | 26.0102150 | -111.3431310 | 333 | 1.2 | 54.1 |
| Viggé-Biaundó (BC) | San Francisco Javier (1744) | 25.8608550 | -111.5435790 | 185 | 17.9 | -46.0 |
| Mulegé (BC) | Sta. Rosalía de Mulegé (1705) | 26.8852500 | -111.9859240 | 340 | 6.8 | 62.4 |
| Comondú (BC) | San José de Comondú (1708) | 26.0592768 | -111.8230614 | 152 | 20.8 | -35.8 |
| Todos Santos (BC) | Ntra. Sra. del Pilar (1748) | 23.4497280 | -110.2255726 | 207 | 1.3 | -53.9 |
| San Ignacio (BC) | San Ignacio de Kadakaamán (1750) | 27.2837780 | -112.8985180 | 289 | 2.1 | 17.8 |
| San José del Cabo (BC) | San José del Cabo Añuití (1730) | 23.0621370 | -109.6955220 | 264 | 3.7 | -4.1 |
| San Luis Gonzaga (BC) | San Luis Gonzaga Chiriyaquí (1740) | 24.9081850 | -111.2908720 | 180 | 6.7 | -58.4 |
| San Francisco de Borja (BC) | San Francisco de Borja Adac (1752) | 28.7442970 | -113.7542010 | 337 | 6.2 | 58.7 |
| San Ignacio (So) | San Ignacio de Cabórica (1770) | 30.6967996 | -110.9243024 | 0 | 2.3 | 61.6 |
| Tubutama (So) | San Pedro y San Pablo Apóstoles | 30.8848383 | -111.4654224 | 0 | 0.9 | 60.0 |
| Atil (So) | San Francisco de Asís (1730) | 30.8443376 | -111.5838051 | 350 | 2.2 | 59.8 |
| Magdalena de Kino (So) | Santa María Magdalena (1688) | 30.6302244 | -110.9730947 | 269 | 1.2 | -0.3 |
| Imuris (So) | Ntra. Sra. del Pilar y Santiago de Cocóspera (1687) | 30.9283656 | -110.6114718 | 0 | 3.1 | 62.2 |
| Oquitoa (So) | San Antonio de Padua (1689) | 30.7434828 | -111.7350092 | 354 | 2.0 | 60.7 |
| Pitiquito (So) | San Diego de Alcalá (1694) | 30.6751626 | -112.0578740 | 126 | 2.4 | -28.9 |
| Tumacácori (Ar) | San José de Tumacácori (1691) | 31.5683752 | -111.0508841 | 352 | 0.6 | 58.1 |
| Río Rico (Ar) | San Cayetano de Calabazas (1756) | 31.4523117 | -110.9593872 | 339 | 1.2 | 53.8 |
| Tucson (Ar) | San Francisco Xavier (del Bac) (1699) | 32.1067079 | -111.0079249 | 357 | 0.6 | 58.4 |
| Caborca (So) | Purísima Concepción de Ntra. Sra. de Caborca (1692) | 30.6975360 | -112.1471278 | 104 | 2.2 | -10.9 |
| Arizpe (So) | Ntra. Sra. de la Asunción (1646) | 30.3367276 | -110.1655977 | 13 | 1.5 | 58.6 |
| El Bosque (Ch) | Sta. Ana de Chinarras (1716) | 28.8127476 | -105.9328097 | 327 | 4.4 | 50.3 |
| San Francisco de Borja (Ch) | San Francisco de Borja Adac (1639) | 27.9026285 | -106.6851962 | 260 | 2.8 | -7.5 |
| Satevó (Ch) | Misión Sto. Ángel Custodio (1764) | 26.9929831 | -107.7144324 | 185 | 5.3 | -57.4 |
| Sinaloa de Leyva (Si) | San Felipe y Santiago (1772) | 25.8218823 | -108.2223910 | 353 | 0.5 | 63.8 |
| Bacubirito (Si) | San Pedro Apóstol (1650) | 25.8080678 | -107.9173146 | 1 | 3.3 | 67.5 |
| Baymena (Si) | Iglesia de Baymena | 26.5238593 | -108.2905519 | 150 | 6.1 | -46.3 |
| Nío (Si) | Ruinas de Nío (1700) | 25.6245484 | -108.3995201 | 117 | 0.5 | -23.9 |
| San Francisco Javier de Satevó (Ch) | San Francisco Javier de Satevó | 27.9547590 | -106.1077980 | 283 | 1.9 | 12.4 |
| Nonoava (Ch) | Ntra. Sra. de Monserrat (1690) | 27.4710673 | -106.7372049 | 248 | 2.0 | -18.4 |
| Santiago Papasquiaro (Du) | Santiago Apóstol | 25.0394287 | -105.4203510 | 358 | 0.7 | 65.6 |
| Sta. Catarina de Tepehuanes (Du) | Sta. Catarina de Tepehuanes | 25.3430313 | -105.7229649 | 270 | 1.9 | 0.8 |
| Villa Ocampo (Du) | San Miguel Arcángel | 26.4398644 | -105.5093186 | 289 | 1.4 | 17.6 |
| San José del Tizonazo (Du) | Sr. de los Guerreros | 25.9642239 | -105.1896614 | 236 | 5.5 | -27.3 |
| Pátzcuaro (Mi) | Iglesia de la Compañía (1747) | 19.5130727 | -101.6073026 | 205 | 3.0 | -56.7 |
| Zacatecas centro | Parroquia de Sto. Domingo (1746) | 22.7762014 | -102.5733650 | 30 | 2.7 | 54.7 |
| San Luis Potosí | Ntra. Sra. de Loreto (1675) | 22.1522431 | -100.9783299 | 353 | 0.0 | 66.8 |
| Parras de la Fuente (Co) | San Ignacio de Loyola (1607) | 25.4370213 | -102.1837422 | 184 | 5.1 | -59.2 |
| La Mesa del Nayar (Na) | Santísima Trinidad de La Mesa (Rey Nayar) (1750) | 22.2153156 | -104.6517626 | 165 | 4.1 | -59.9 |
| Jesús María (Na) | Iglesia de Jesús, María y José (1745) | 22.2508230 | -104.5173572 | 101 | 7.1 | -7.4 |
| Guanajuato | Templo de la Compañía de Jesús (1747) | 21.0170349 | -101.2526567 | 14 | 4.2 | 68.4 |